\documentclass[runningheads]{llncs}
\usepackage[T1]{fontenc}
\usepackage{amsmath}
\usepackage{amsfonts}
\usepackage{amssymb}
\usepackage{bigints}
\usepackage{subcaption}
\usepackage{mathtools}
\usepackage{graphicx}
\begin{document}
\title{Generating Elementary Integrable Expressions}
%
%
\author{Rashid Barket\inst{1}\orcidID{0000-0002-9104-4281} \and 
Matthew England\inst{1}\orcidID{0000-0001-5729-3420} \and
J\"{u}rgen Gerhard\inst{2}}
\authorrunning{Barket et al.}
%
\institute{
Coventry University, Coventry, United Kingdom\\
\email{\{barketr,matthew.england\}@coventry.ac.uk} \and 
Maplesoft, Waterloo, Canada\\
\email{jgerhard@maplesoft.com}}

\maketitle              
\begin{abstract}
There has been an increasing number of applications of machine learning to the field of Computer Algebra in recent years, including to the prominent sub-field of Symbolic Integration. However, machine learning models require an abundance of data for them to be successful and there exist few benchmarks on the scale required. While methods to generate new data already exist, they are flawed in several ways which may lead to bias in machine learning models trained upon them.  In this paper, we describe how to use the Risch Algorithm for symbolic integration to create a dataset of elementary integrable expressions.  Further, we show that data generated this way alleviates some of the flaws found in earlier methods.    

\end{abstract}
\section{Introduction}

\subsection{Machine Learning and Computer Algebra}

A key feature of a Computer Algebra System (CAS) is its exactness: when prompted for a calculation, a CAS is expected to return the exact answer (or no answer if the calculation is not feasible), as opposed to an approximation to an answer. Due to this restraint, it seems as though Machine Learning (ML) and Computer Algebra do not work well together due to the probabilistic nature of ML: no matter how well-trained an ML model is, it can never guarantee perfect predictions. However, rather than trying to use ML to predict a calculation in place of a CAS, we can instead use ML in conjunction with a CAS to help optimize and/or select the symbolic computation algorithms implemented within.  Such a combination of ML and symbolic computation preserves the unique selling point of a CAS. The earliest examples of such ML for CAS optimisation known to the authors are: 
Hunag et al. \cite{Huang2014} which used a support vector machine to choose the variable ordering for cylindrical algebraic decomposition; and Kuipers et al. \cite{Kuipers2015} which used a Monte-Carlo tree search to find the representation of polynomials that are most efficient to evaluate.

\subsection{Symbolic Integration Meta-Algorithms}

Our interest is the integrate function of a CAS, which takes an integrand and produces an integral (either definite or indefinite). In most CASs, and certainly in Maple where the authors focus their work, the integrate function is essentially a meta-algorithm: it accepts a mathematical expression as an input, does some pre-processing on the expression, and then passes the processed problem to one of a selection of available sub-algorithms.  In Maple, the function will try a list of such sub-algorithms in turn until one is found that can integrate the expression, in some cases first querying a guard as to whether that sub-algorithm is applicable to the input in question. If none of these methods work, the function simply returns the input back as an unevaluated integral (implying that Maple cannot integrate it). 

Currently, as of Maple 2023, these sub-algorithms for \texttt{int} are tried in the same pre-set order for every input, and outputs the answer of the first sub-algorithm that works. There are currently 11 sub-algorithms to choose from. The list of sub-algorithms is available on the Maple help page\footnote{\url{www.maplesoft.com/support/help/maple/view.aspx?path=int\%2fmethods+}} for the function.

The first use of ML is to improve the integrate function's efficiency. A similar approach was taken by Simpson et al. \cite{Simpson2016} for the resultant function (see Definition \ref{def:resultant} later). After applying a neural network to classify which algorithm (of four possible choices) to use, the authors test their model on a random sample of several thousand inputs. Maple's existing meta-algorithm took 37,783 seconds to finish its computations, whereas the sub-algorithm choices from neural network took only 12,097 seconds-a significant improvement with a 68\% decrease in runtime. There were also gains against Mathematica with a 49\% decrease in runtime. We hope to achieve similar results with the integrate function.

\begin{figure}[t]
    \centering
    \includegraphics[width=.75\textwidth]{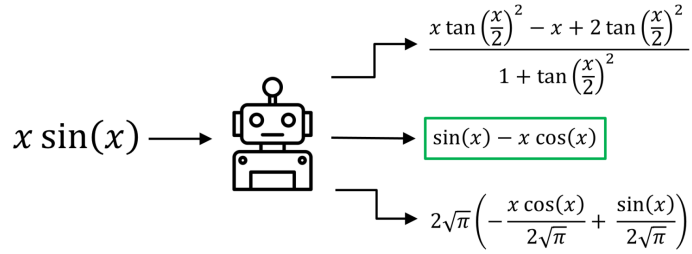}
    \caption{The output of $\int  x\sin(x)$ from each successful sub-algorithm. The main output chosen in this case is the shortest expression chosen by an ML model, from sub-algorithm 2.}
    \label{fig:Ouput}
\end{figure}

The second motivation to use ML is in optimizing the output. To gain a better understanding of this, consider what happens in Figure \ref{fig:Ouput} when you integrate the function $f(x) = x\sin(x)$ in Maple and ask it to try all possible sub-algorithms. When $f(x)$ is integrated, there are three successful outputs that come from three different sub-algorithms. Each output is expressed differently but are all mathematically correct and equivalent. We wish to choose the simplest output, which in this case is $\int f(x) =~\sin(x) - x\cos(x)$.

\subsection{Motivation}

The goal of the data generation method described in this paper is to be able to produce many integrable expressions to train a ML model on. There is not enough benchmark/real-world data to train a model on, hence why these data generation methods are needed. There does currently exists data generation methods. Lample \& Charton \cite{Lample2020} produce three methods for developing integrable expressions: FWD, BWD, and IBP (described in detail in Section \ref{sec:LampleCharton}). These methods have drawbacks which the data generation method we propose will handle.

The FWD method, which generates a random expression and calculates its integral, tends to produce short integrands and long integrals. Furthermore, the FWD method will typically not have an elementary integral. This is especially evident for longer randomly generated expressions and/or expressions with denominators. This means the FWD method will take numerous attempts before finding a valid (integrand, integral) pair. The BWD method, which generates an expression and calculates its derivative, has the opposite problem of long integrands and short integrals. The IBP, or integration by parts method, produces expressions that are too similar (meaning that the expressions only differ by their coefficients) which is discussed in Section~\ref{sec:LampleCharton}. Hence, a dataset of (integrand, integral) pairs is needed for this method to work.

We propose generating (integrand, integral) pairs based on the Risch Algorithm. For one, the method will always produce an elementary integrable expression, something FWD cannot guarantee. This data generation method also does not have the issue of varied lengths between the integrands and integrals because of various parameters available from the data generation method, alleviating the length issues in the FWD and BWD methods. Lastly, this method does not require a dataset of known integrals and also does not produce expressions too similar to the rest of the dataset, which IBP suffers from. Data generation based on the Risch algorithm produces a variety of non-trivial, unique expressions that current data generation methods do not offer. Further discussion of current methods and the new method presented are discussed in Sections \ref{sec:LampleCharton} and \ref{sec:Discussion}.

\subsection{Contributions and Plan}

This paper will focus on how to generate sufficient data to make our planned application of ML.  In Section \ref{sec:current_methods}, we overview the existing methods of data generation for the problem that we found in the literature, explaining why they are not suitable alone for our needs.  Then in Section \ref{sec:Risch}, we review the classical Risch algorithm which will be the basis of our new data generation method introduced in Section \ref{sec:gen_expr} which identifies constructive conditions for an integrand to be elementary integrable.  We finish in Section \ref{sec:Discussion} with a discussion on the advantages of this approach over the existing methods and what future steps still need to be undertaken.

\section{Existing Datasets and Data Generation Methods}\label{sec:current_methods}

An important aspect of a successful ML model is that it is generalisable. That is, the model should perform well on all inputs it receives and not just inputs that look very similar to the training data. There are existing datasets and data generation methods for symbolic integration. However, each comes with its own sets of limitations that prevent an ML model trained on them to generalise well on all real-world data.

\subsection{Deep Learning For Symbolic Mathematics}\label{sec:LampleCharton}

In their paper (with the same name of this subsection), Lample and Charton \cite{Lample2020} experiment on using deep learning to perform the tasks of symbolic integration and solving ordinary differential equations directly. To achieve this, they used a seq2seq model $-$ a neural network architecture used in natural language processing for mapping sequences of tokens (usually words to another such sequence) $-$ in the form of a transformer\footnote{the same model which is the basis for ChatGPT}. 

There are different classes of integrals that can be output based on its complexity.

\begin{definition}[Elementary Function]\label{def:elementary}
    A function that is defined as the sum, product, root, or composition of finitely many polynomial, rational, trigonometric, hyperbolic, and exponential functions (and their inverses) is considered elementary.
\end{definition}

An expression that, when integrated, produces an elementary function is said to be \textit{elementary integrable}. Most expressions one encounters in a first-year calculus class will be elementary integrable. An example of an expression that is not elementary integrable is $f(x) = \frac{1}{\log{x}}$. When $f(x)$ is integrated, the result usually produced is $\textrm{li}(x)$, a non-elementary function known as the Logarithmic Integral special function\footnote{\url{https://dlmf.nist.gov/6.2}}.

\begin{figure}[b]
    \centering
    \includegraphics[width=.70\textwidth]{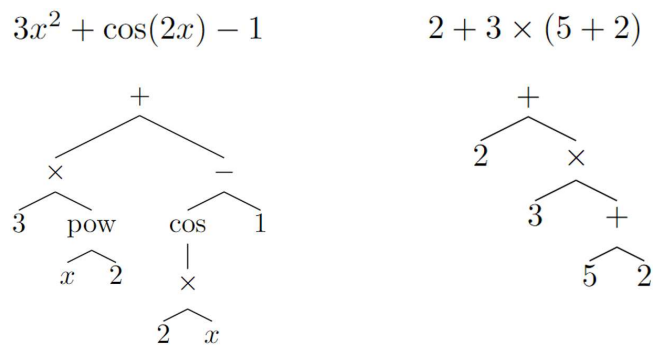}
    \caption{Tree representation for $3x^2 + cos(2x) -1$ and $2 + 3 \times (5+2)$ from \cite{Lample2020}. With some restrictions as to how the trees are constructed, there is a one-to-one mapping between an expression and its tree.}
    \label{fig:expr_tree}
\end{figure}

The authors of \cite{Lample2020} created a novel way of generating data to train a transformer. Expressions are viewed as trees, where the internal nodes are operators or function names ($+$, $sin$, etc.), and the leaves are constants and variables as exemplified in Figure \ref{fig:expr_tree}. An algorithm is developed to generate trees of varying length so that these expressions can be used for training the model. They added structure to the trees in the form of restriction on internal nodes and leaves such that every random tree created is a valid symbolic expression.

They treated this as a supervised learning problem, generated the following three methods to take such symbolic expressions and produced labelled training pairs:
\begin{itemize}
    \item FWD: Integrate an expression $f$ through a CAS to get $F$ and add the pair ($f, F$) to the dataset.
    \item BWD: Differentiate an expression $f$ to get $f'$ and add the pair ($f', f$) to the dataset.
    \item IBP: Given two expressions $f$ and $g$, calculate $f'$ and $g'$. If $\int f'g$ is known then the following holds (integration-by-parts):
    \[
    \int fg' = fg - \int f'g.
    \]
    Thus we add the pair ($fg'$, $fg - \int f'g$) to the dataset. 
\end{itemize}

While these three methods can generate plenty of elementary integrable expressions, they come with many limitations that can cause an ML model to overfit on the training data. For both the FWD and BWD methods, they tend to create expressions with patterns in the length. For FWD, the integrand is on average shorter than the resulting integral. BWD suffers from the opposite problem: long integrands and short integrals. Individually, these cause problems when training the transformer as the model is fitted too closely to these patterns, leading to overfitting. For example, the results from Lample \& Charton show that when a model is trained on only FWD data and tested on BWD data, it only achieves an accuracy of 17.2\%, and similar results are shown for training on BWD and testing on FWD. They of course train the model on all three data generation methods, but it is not clear if this addresses all the overfitting or simply encodes both sets of patterns.

Furthermore, these data generation methods suffer from producing expressions that are far too similar between the training and testing data. Piotrowski et al. \cite{Piotrowski2019} perform a simple analysis of substituting all coefficients with a symbolic \texttt{CONST} token. They examine how many expressions show up in the training set that are also the same in the testing set modulo constant and sign. For the FWD, BWD, and IBP methods, the percentage of unique data were 35\%, 75\% and 24\%, respectively. A key principle of machine learning is that the testing data should be independent of the training data but this casts doubt on whether this is possible through the partition of a dataset containing such similar examples.  This may be considered an example of ML ``\emph{data leakage}''. Data leakage is a significant issue in machine learning. It happens when the training data we use contains the information that the model is trying to predict. This can result in unpredictable and poor predictions once the model is deployed.

\subsection{Other Existing Datasets}

Currently, there are not that many (public) benchmark datasets in the field of symbolic integration, or indeed Computer Algebra more broadly. Maplesoft has an in-house test suite of integrable functions that they use to ensure software quality is maintained when making changes to \texttt{int}. There are 47,745 examples in the Maple test suite. Of these, only 8,174 had elementary integrands with elementary integrals which we currently study. We provide some information from the remaining (integrand, integral) pairs in Table \ref{table:MapleStats}. 

\begin{table}[h]
\centering
\begin{tabular}{l|c|c}
                                                        & \multicolumn{1}{l|}{Integrand} & \multicolumn{1}{l}{Integral} \\ \hline
\multicolumn{1}{l|}{Average Number of Operands}         & 2.59                           & 6.52                          \\ \hline
\multicolumn{1}{l|}{Largest Number of Operands}         & 16                             & 300                           \\ \hline
\multicolumn{1}{l|}{Is a Polynomial}                    & 996                            & 1221                          \\ \hline
\multicolumn{1}{l|}{Average Polynomial Degree}          & 1.80                            & 2.79                          \\ \hline
\multicolumn{1}{l|}{Largest Polynomial Degree}          & 199                            & 200                          \\ \hline
\multicolumn{1}{l|}{Contains Exponentials}              & 932                            & 1072                          \\ \hline
\multicolumn{1}{l|}{Contains Logarithms}                & 756                            & 3136                          \\ \hline
\multicolumn{1}{l|}{Contains Trig or Arctrig functions} & 2080                           & 2512                          \\ \hline
\multicolumn{1}{l|}{Contains Radicals}                  & 2024                           & 2274                          \\ \hline
\multicolumn{1}{l|}{Contains Complex Numbers}           & 558                            & 685                           
\end{tabular}
\caption{A summary of the (integrand, integral) pairs in the Maple test suite (total 8174). We only kept functions with elementary integrands which had elementary integrals}
\label{table:MapleStats}
\end{table}

These number of examples would not be sufficient to train a deep learning model; for reference, Lample and Charton \cite{Lample2020} have access to 88 million examples in \textit{Deep Learning for symbolic Mathematics}. One great property about the Maple dataset is that it was partly developed as a continuous response to feature requests and bug reports that users would make when using \texttt{int} in Maple. Thus, it can be said to represent the range of examples of interest to Maple users. Using this dataset to evaluate any models trained would help provide evidence that the model generalizes well for our planned use.

Rich et al. \cite{Rich2018} developed a Rule-Based Integrator, more commonly known as RUBI. RUBI integrates an expression by applying a collection of symbolic integration rules in a systematic way. Along with RUBI, the authors have compiled a dataset of 72,000 integration problems. There are 9 different main categories of functions that exist in the dataset with many examples coming from various textbooks and papers. Similar to the Maple test suite, this dataset would be good for evaluating a model but due to the size of the dataset, it would not be sufficient for training a model, at least not a deep learning based model.  We thus use the rest of our paper describing a new method.

\section{The Risch Algorithm} \label{sec:Risch}

The data generation method in this paper is based on the Risch algorithm. To explain the entire Risch algorithm would need us to introduce a lot of theory before even getting to the algorithm explanation. Instead, we will focus on the key parts of the algorithm to help the reader get an intuitive understanding of how it works and refer to \cite{Bronstein2005} or \cite[Ch. 11, 12]{Geddes1992} for a more detailed explanation. 

For the Risch algorithm to work, we allow elementary extensions over a differential field $K$. A differential field is a field with the derivative operator $D$ such that $D(a+b) = D(a) + D(b)$ and $D(ab)= aD(b) + bD(a)$. A constant $c$ is defined as $Dc=0$. We usually write the derivative $Da=a'$. 

Let $G$ be an extension field of a differential field $F$. For an element $\theta \in G$, We say that $G$ is an elementary extension of $F$ if $\theta$ is one of the following:
\begin{enumerate}
    \item \textbf{logarithmic}: $\theta$ = $\log(u)$, $u \in F$.
    \item \textbf{exponential}: $\theta$ = $e^u$, $u \in F$.
    \item \textbf{algebraic}: $\exists p \in F$ such that $p(\theta)=0$.
\end{enumerate}

An arbitrary amount of extensions are allowed. Rather than using $G$ to represent the extension, we instead denote $F_{n-1} = K(\theta_1, \cdots, \theta_{n-1})$ as the previous differential field and $F_n = F_{n-1}(\theta_n)$ as the current elementary extension. Typically, we have $K = \mathbb{Q}(x)$ as the base differential field.

This paper will focus solely on logarithmic and exponential extensions. We now introduce Liouville's theorem that states exactly what the form of the integral will be, if it exists.

\begin{theorem}[Liouville's Theorem: Thm 5.5.1 in \cite{Bronstein2005}]\label{theorem:liouville}
    Let $K$ be a differential field and $f \in K$. Let $E$ be an elementary extension of $K$. If $\int f \in E$ exists, then there are $v_0,\cdots,v_m \in K$ and constants $c_0, \cdots, c_m \in K$ such that

\[
    \int f = v_0 + \sum_{i=0}^{m} c_{i}\log(v_i)
\]
    
\end{theorem}

Liouville's Theorem gives an explicit representation for the integral of $f$ if it is elementary integrable. The Risch algorithm and the subsequent algorithms for computing an integral are based on Liouville's Theorem. The Risch algorithm will divide the input into two different parts. Then, the integral for both parts will take the form of Theorem \ref{theorem:liouville}.    

\subsubsection{Risch Algorithm (Chapter 12 in \cite{Geddes1992}):}
Let $F_{n} = F_{n-1}(\theta_n)$ be a differential field of characteristic 0 where $\theta_{n}$ is elementary over $F_{n-1}$, and $\theta_{i}' \neq 0, 1 \leq i \leq n$. For any rational function $f=g/b$ with respect to $\theta_n$, you can divide the numerator with remainder $g=Pb+R$ where $\deg_{\theta_{n}}(R) < \deg_{\theta_{n}}(b)$, and have $f=P+\frac{R}{b}$. If $f$ is elementary integrable, it follows that $\int f=\int P+ \int \frac{R}{b}$. We call $P$ the polynomial part and $\frac{R}{b}$ the rational part. We study these two parts for the rest of the section and then develop ways to generate elementary integrable expressions from both these parts in Section \ref{sec:gen_expr}. 

\subsection{The Rational Part}

Suppose we wish to integrate $\frac{R}{b}$, $R,b \in F=K(x)(\theta_1,\cdots, \theta_n)$. There are two algorithms used to compute this integral: Hermite Reduction and the Trager-Rothstein (TR) method. Which algorithm is used depends on whether the denominator $b$ is square-free or not.

\begin{definition}[Square-free]\label{def:DF}
We say $a \in K[x]$ is square-free if $a$ has no repeated factors i.e. $\nexists b \in K[x]$ such that $\deg(b) > 0$ and $b^2 | a$. Equivalently, $\gcd(a,a')=1$
\end{definition}

When our denominator is not square-free, we use Hermite Reduction.

\begin{theorem}[Hermite Reduction: Thm 5.3.1 in \cite{Bronstein2005}]\label{theorem:Hermite}
    Suppose we want to integrate $\int \frac{R}{b}$, where $R$,$b \in F[\theta]$ and $\deg_{\theta}(R) < \deg_{\theta}(b$). Use the square-free factorization $b=b_{1}b_{2}^2 \cdots b_{k}^k$ where $b_i$ is square-free. Let $T = b/b_k^k$. Let $\sigma$ and $\tau$ be the solutions to the diophantine equation
    \begin{equation*}
        \sigma b_{k}'T + \tau b_k = R.
    \end{equation*}
    Then Hermite reduction tells us that
    \begin{equation*}
        \int \frac{R}{b} = \frac{-\sigma(k-1)}{b_k^{k-1}} + \int \frac{\tau + \frac{\sigma'}{k-1}T}{\frac{b}{b_k}}.
    \end{equation*}
\end{theorem}

The main part to notice is that the resulting integral on the right hand side of the equation has a denominator that is at least one degree less than the input denominator (because we divide $b$ with its highest degree factor $b_k$). This algorithm is used recursively until the resulting integral's denominator has degree one, allowing us to conclude that it is square-free. When this point is reached then the TR-method is used on the remaining integral.  This method makes use of the following tool from computational algebra.

\begin{definition}[Resultant]\label{def:resultant} Suppose we have the following two polynomials with roots $\alpha_i$ and $\beta_j, \alpha_m \neq 0 \neq \beta_n$:
\begin{align*}
A &= a_0 + \cdots a_{m}x^m = a_m\prod\limits_{i=1}^m (x-\alpha_i) \\
B &= b_0 + \cdots b_{n}x^n = b_n\prod\limits_{j=1}^n (x-\beta_j) 
\end{align*}
Then their resultant is defined as $res_x(A,B) = (-1)^{mn}b_n^{m}a_m^{n}\prod\limits_{j=1}^n \prod\limits_{i=1}^m (\beta_j - \alpha_i)$

This implies that 
\begin{enumerate}
    \item $res(A,B) = \pm res(B,A)$
    \item $res(A, BC) = res(A,B)res(A,C)$ 
\end{enumerate}
for all nonzero polynomials $A,B,C$.

\end{definition}
Note that the resultant can be calculated without finding the roots of each polynomial by using Sylvester's Matrix described on page 285 of \cite{Geddes1992}. 

Given an integral with square free denominator $\int \frac{R}{b}$, we define the Trager-Rothstein resultant polynomial (TR-resultant) as res$_{\theta}(R-zb',b)$. We will forego the details of the rest of the algorithm and focus on a key theorem involving the TR-resultant polynomial.

\begin{theorem}[Thm 12.7 in \cite{Geddes1992}]\label{theorem:TR}
    Suppose we are integrating $\int \frac{R(x)}{b(x)}$, where $R(x)$, $b(x) \in F[x]$ and $b(x)$ is square-free. Then we have that $\int \frac{R(x)}{b(x)}$ is elementary integrable if and only if all the roots in $z$ of the TR-resultant are constants.
\end{theorem}

Theorem \ref{theorem:TR} is the key theorem that tells us whether a rational expression will be elementary integrable or not, either in application to itself if the denominator is square free, or in application to the final integral from Hermite reduction if not. This theorem will also be the key theorem to create the data generation method for rational expressions.

\subsection{The Polynomial Part}

Suppose we are integrating $P$, a polynomial in $F[\theta]$. We again only focus on logarithmic and exponential extensions from our field. There are two different procedures to integrate $P$ based on if the extension is logarithmic or exponential. 

\subsubsection{Logarithmic extension:} Let $P=p_0 + p_1\theta + \cdots + p_l\theta^m$ where $\theta=\log(u), u, p_i \in F_{n-1}$. It can then be shown that
\begin{equation}\label{eqn:polyPart_log}
    \int p_0 + \cdots + p_m\theta^m = q_0 + \cdots + q_{m+1}\theta^{m+1} + \sum_{i=1}^k c_i\log(v_i),
\end{equation}
where $q_{m+1} \in K, q_i \in F_{n-1} (1 \leq i \leq m), c_j \in K, v_j \in F_{n-1} (1 \leq j \leq k)$. The idea behind integrating $P$ is to differentiate Equation (\ref{eqn:polyPart_log}) and then equate the coefficients of like powers of $\theta$ to solve for each $q_i$. The details of this can be found in \cite[page 540]{Geddes1992}.

\subsubsection{Exponential extension:} The exponential case is similar to the logarithmic case, however a couple of adjustments need to be made. The first adjustment is that polynomial exponents are allowed to be negative for exponential extensions. Thus, $P = p_{-l}\theta^{-l} + \cdots + p_0 + \cdots + p_m\theta^m$ and Equation (\ref{eqn:polyPart_log}) becomes:
\begin{equation}\label{eqn:polyPart_exp}
    \int p_{-l}\theta^{-l} + \cdots + p_0 + \cdots + p_m\theta^m = q_{-l}\theta^{-l} + \cdots + q_0 + \cdots + q_{m}\theta^{m} + \sum_{i=1}^k c_i\log(v_i).
\end{equation}

Note that in Equation (\ref{eqn:polyPart_exp}), the answer has a highest degree of $m$ instead of $m+1$. The steps for equating like powers of $\theta$ differ between Equation (\ref{eqn:polyPart_log}) and (\ref{eqn:polyPart_exp}), and we will see an example of this difference soon in Section \ref{sec:poly_gen}.

\section{Data Generation based on the Risch Algorithm}\label{sec:gen_expr}

In order to generate elementary integrable expressions, we will do what the Risch algorithm does as an initial step: generate polynomial expressions and rational expressions separately. Polynomial expressions and rational expressions can then be combined together through the additive property of integrals. We first focus our attention on the simpler case: the polynomial part. Then, we will show how to generate rational expressions.

\subsection{Polynomial Integrable Expressions}\label{sec:poly_gen}

Generating polynomial expressions (in $\theta$) that are elementary integrable requires choosing the coefficients $q_i$ from Equation (\ref{eqn:polyPart_log}) or (\ref{eqn:polyPart_exp}) ourselves. We differentiate the equation and equate coefficients of like powers of $\theta$, resulting in a system of differential equations. The randomly chosen $q_i$'s are substituted into this system to generate the integrable expression. 

It turns out that this is no better than just using the BWD method, i.e., we select a random polynomial in $\theta$ with random coefficients in $F_{n-1}$ and take its derivative. This is not as general as it could be; one would also have to generate a random integrable expression in the smaller field $F_{n-1}$. For the sake of simplicity, we omit this step here, which could be done recursively or by using the BWD method. We provide a small example of the BWD method for polynomials in $\theta$ to show how the data is generated.

\begin{example}
    Suppose we want to generate a degree 2 polynomial in $\mathbb{Q}(x)[\theta]$ where $\theta=\ln(\frac{1}{x})$. The coefficients in $\theta$ must be in the previous field $\mathbb{Q}(x)$. For simplicity, the logarithms in Equation (\ref{eqn:polyPart_log}) are omitted. The following coefficients are generated randomly:
    \begin{itemize}
        \item $q_0 = -7 + 8x + \frac{2}{x}$
        \item $q_1 = -5 + 4x - \frac{6}{x}$
        \item $q_2 = 1+2x$
    \end{itemize}
which results in the polynomial 
\[
P = \left(1+2 x \right) \ln \! \left(\frac{1}{x}\right)^{2}+\left(-5+4 x -\frac{6}{x}\right) \ln \! \left(\frac{1}{x}\right)-7+8 x +\frac{2}{x}.
\]

When differentiated, we get

\[
P' = 2 \ln \! \left(\frac{1}{x}\right)^{2}+\left(-\frac{2 \left(1+2 x \right)}{x}+4+\frac{6}{x^{2}}\right) \ln \! \left(\frac{1}{x}\right)-\frac{-5+4 x -\frac{6}{x}}{x}+8-\frac{2}{x^{2}}
\]
and the pair ($P', P$) is added to our dataset.
\end{example}

\subsection{Rational Integrable Expressions}\label{sec:rational_gen}

As we will see in a moment, generating rational integrable expressions is more complex than the polynomial case. We will introduce some strategies to generate integrable expressions with square-free denominators (using the TR-method) as well as non square-free denominators (using a combination of Hermite reduction and the TR-method). Note that most of the examples shown here will be using the extension $\theta=\log(u)$ as this is the harder case to solve. However, extensions with $\theta=e^u$ will also appear in the dataset produced.

\subsubsection{Square-Free Denominators:}\label{sec:TR_gen}

In the normal use of the TR-method, the input is a rational elementary function $\frac{R}{b}$ such that $\deg_\theta(R) < \deg_\theta(b)$ and $b$ is square-free. The method then outputs the elementary integral of $\frac{R}{b}$, or fails if Theorem~\ref{theorem:TR} does not hold. Our goal is to discover polynomials $R, b \in F[\theta]$ such that $\frac{R}{b}$ is guaranteed to be elementary integrable. The main idea behind the process is to fulfill the conditions of Theorem \ref{theorem:TR} so that we know for sure that the expression is elementary integrable. To accomplish this, the general outline is as follows.
\begin{enumerate}
    \item Randomly generate the denominator $b$ in its square-free factorization, and keep that fixed.
    \item Create a partial fraction decomposition where the denominators are all factors of $b$, and the numerators are polynomials in $\theta$ of degree 1 less than the denominator, with symbolic coefficients.
    \item Compute the TR-resultant.
    \item The symbolic coefficients of $R$ must be chosen in a way that ensures the roots of the resultant are constant.
    \begin{enumerate}
        \item If the resultant only has factors of degree 2 or less, solve directly for the roots and set each root equal to a constant.
        \item Otherwise, the resultant has irreducible factors of degree 3 or higher. Divide the resultant by the leading coefficient to make it monic. Then, the symbolic coefficients must be chosen in such a way that each coefficient of this is constant.
    \end{enumerate}      
\end{enumerate}

We first put our input into partial fraction form with symbolic coefficients because when the resultant is calculated, the TR-resultant factors in a way similar to how $b$ factors (See Definition \ref{def:resultant}). We can see this with the following example.

\begin{example}\label{eg:rational_easy}
    
    Let $b=\theta^4 - 2\theta^2 - 2\theta^3 - 2\theta - 3$ where $\theta=\log(x)$, $F=\mathbb{Q}(x)(\log(x))$ and we have only done a single extension so $n=1$. We wish to discover a class of numerators $R$ so that $\frac{R}{b}$ integrates.
    \begin{itemize}
        \item Note that $b$ factors into $b=(\theta + 1)(\theta - 3)(\theta^2 + 1)$.
        \item We create the partial fraction representation of our input:
        \begin{center}
            $\frac{a \! \left(x \right)}{\theta +1}+\frac{b \! \left(x \right)}{\theta -3}+\frac{c(x)\theta + d(x)}{\theta^{2}+1}$,
        \end{center}
        where $a,b,c,d \in F_{n-1}=\mathbb{Q}(x)$.
        \item The factored form of the TR-resultant of $\frac{R}{b}$ is
        \begin{center}
            $-(a(x)x - z)(b(x)x - z)(c(x)^2x^2 - 4c(x)xz + d(x)^2x - 2d(x)xz + xz^2 + 4z^2)$.
        \end{center}
        \item Recall that by Theorem~\ref{theorem:TR}, we need the roots of the resultant to be constant. Setting each factor of the resultant equal to a constant and solving for the symbolic coefficients, we get that $a(x)=\frac{C_1}{x}, b(x)=\frac{C_2}{x}, c(x)=\frac{C_3}{x}$, and $d(x)=\frac{C_4}{x}$  for any $C_1, C_2, C_3, C_4 \in \mathbb{Q}$.
        \item Therefore, $\frac{R}{b} = \frac{C_{1}}{x \left(\theta +1\right)}+\frac{C_{2}}{x \left(\theta -3\right)}+\frac{C_{3}\theta+C_{4}}{x(\theta^{2}+1)}$ is elementary integrable for any choice of those constants. We find that:
        \begin{align*}
            \int \frac{R}{b} & =\frac{C_{3} \log \! \left(\log \! \left(x \right)^{2}+1\right)}{2}+C_{4} \arctan \! \left(\log \! \left(x \right)\right) \\
            &\qquad +C_{1} \log \! \left(\log \! \left(x \right)+1\right)+C_{2} \log \! \left(\log \! \left(x \right)-3\right).
        \end{align*}
    \end{itemize}
\end{example}

In Example \ref{eg:rational_easy}, take note that the factored form of the resultant is similar to the factored form of the denominator $b$: that is, the degree in $z$ of each factor of the resultant is the same as the degree in $\theta$ of each factor of $b$. As well, each symbolic coefficient in the numerator of each partial fraction were also the same unknowns that show up in each factor of the resultant.

Example \ref{eg:rational_easy} only had linear and quadratic irreducible factors. These are quite easy to solve by just isolating the unknown or using the quadratic formula. In general, degree 3 and higher irreducible factors in the resultant will be much harder to solve. Trying to solve for the roots of an irreducible degree 3 resultant means using the Cardano formula, which produces huge answers for the root.  We find that when trying to equate any of the roots to a constant and solving for the conditions of $R$ like in Example \ref{eg:rational_easy}, the expression size blows up and the solution starts to involve many radicals. Since radicals do not lie within our field, the symbolic coefficients then need to be chosen in a way such that the radicals disappear which adds an extra layer of complexity.  The formulae size is even worse in degree 4 and then there is not even any such formula in surds for higher degree.  So instead when the resultant has factors of degree higher than two, we look at two alternative options:  assume the numerator to be of a specific form or analyse the resultant qualitatively to figure out the conditions of the numerator. We show the former with the following example.

\begin{example}\label{eg:rational_hard}
    Suppose $\theta=\ln(x)$, $F=\mathbb{Q}(x)(\ln(x))$ and $b=x(\theta^3 - x)$. Note that $b$ is square-free in $F$. The first step is to create a partial fraction decomposition with denominator $b$ and symbolic coefficients for the numerator. Let 
    \[
    \frac{R}{b} = \frac{a(x)\theta^2 + b(x)\theta + c(x)}{x(\theta^3 - x)}.
    \]
    The TR-resultant is computed as
    \begin{align*}
        &\left(-x^{3}-27 x^{2}\right) z^{3}+\left(27 x^{2} a \! \left(x \right)+9 x^{2} b \! \left(x \right)+3 x^{2} c \! \left(x \right)\right) z^{2}\\
        &\qquad +\left(-9 x^{2} a \! \left(x \right)^{2}-3 x^{2} a \! \left(x \right) b \! \left(x \right)-9 x b \! \left(x \right) c \! \left(x \right)-3 x c \! \left(x \right)^{2}\right) z \\
        &\qquad +a \! \left(x \right)^{3} x^{2}+3 a \! \left(x \right) b \! \left(x \right) c \! \left(x \right) x -b \! \left(x \right)^{3} x +c \! \left(x \right)^{3}.
    \end{align*}

    Finding the solution to the roots explicitly produces huge expressions for $a(x), b(x)$ and $c(x)$ and involve radicals outside our field. Instead, we assume the form of the symbolic coefficients to find a set of solutions. We will assume they are quadratic polynomials (an arbitrary choice). Let 
    \begin{itemize}
        \item[$\bullet$] $a(x)=a_{2}x^2 + a_{1}x + a_0$, 
        \item[$\bullet$] $b(x)=b_{2}x^2 + b_{1}x + b_0$, 
        \item[$\bullet$] $c(x)=c_{2}x^2 + c_{1}x + c_0$, 
    \end{itemize}
    for $a_i, b_i, c_i \in \mathbb{Q}, 0 \leq i \leq 2$. Since the resultant is cubic in $z$, it will have three roots. First, substitute the assumed form of the three coefficients into the resultant. Note the leading coefficient of the resultant is $(-x^{3}-27 x^{2})$. Then, let our resultant be equal to 
    \[
    (-x^{3}-27 x^{2})(z-r_0)(z-r_1)(z-r_2), r_1, r_2, r_3 \in \mathbb{Q}.
    \]
    Consider the equation formed by setting the TR-resultant computed earlier equal to the form just above.  Let us move the terms to one side so we have an expression equal to 0. We may now solve for each coefficient of $z$ to be 0 giving the following solution
    \begin{align*}
        &\{\mathit{a_0} = 3 \mathit{c_1}, \mathit{a_1} = 0, \mathit{a_2} = 0, 
        \mathit{b_0} = 0, \mathit{b_1} = 0, \mathit{b_2} = 0, \\
        &\qquad \mathit{c_0}=0, 
        \mathit{c_1}~=~\mathit{c_1}, \mathit{c_2}~=~0, \mathit{r_1} = \mathit{c_1}, \mathit{r_2} = \mathit{c_1}, \mathit{r_3} = \mathit{c_1}\}.
    \end{align*}
    This can now be substituted into $R$ to produce
    \begin{center}
        $\int \frac{R}{b} = \int \frac{3c_{1}\ln(x)^2 + c_{1}x}{x(\ln(x)^3 + x} = \mathit{c_1} \ln \! \left(\ln \! \left(x \right)^{3}+x \right)$.
    \end{center}

\end{example}

In Example \ref{eg:rational_hard}, we assumed a particular form for the symbolic coefficients to find a solution. This is a quick way to find a set of solutions, however this does not mean we have found all the solutions like with the linear and quadratic cases. Instead, we should try to fulfill the conditions of 4(b). That is, the symbolic coefficients are chosen in a way such that all of the coefficients of the TR-resultant are constant. To see why this is true, we give an informal proof.

Let the TR-resultant be $f \in K[z]$. We can assume $f$ is monic because if it were not, we will divide out the leading coefficient from the resultant to make $f$ monic. Let $F$ be the algebraic closure of $K(x)$, so that $f \in F[z]$. Factor $f$ over $F$ to get $f = \prod_i (z - a_i), a_i \in F$. Each $a_i$ is a root of $f$. If we want the roots $a_i$ to be constant, they should belong to the algebraic closure $\Bar{K}$. In that case, the coefficients of $f$ should also belong to $\Bar{K}$ because they are the polynomials of $a_i$, and they belong to $K[x]$ because of how we defined $f$. Thus, they belong to $\Bar{K} \cap K[x]$ which is $K$. Therefore, $f$ must have constant coefficients for the roots to be constant.

\subsubsection{Non Square-Free Denominators:}\label{sec:hermite_gen}

When computing the elementary integral of a rational function $\frac{R}{b}$, the first step is to check whether $b$ is square free or not. Similarly, what technique used to generate an elementary integrable expression depends on whether the fixed denominator $b$ starts as square-free or not. Let us now assume $b$ is not square-free, so the TR-method cannot be used currently. We first set up the problem just as with the square-free case: put $b$ in partial fraction form and set symbolic coefficients for each partial fraction. The difference is that before, we would invoke the TR-method. However, $b$ is not square free yet. Thus, we use Theorem \ref{theorem:Hermite}, Hermite Reduction, recursively until we get a resulting integral whose denominator is square-free. Then, we use Theorem \ref{theorem:TR} just as before to find the conditions on $R$ that make the whole expression $\frac{R}{b}$ elementary integrable. The main benefit of non square-free denominators is that there will be more choices of freedom in choosing the symbolic coefficients compared to the square-free case. This is shown with the example below.

\begin{example}\label{eg:hermite}
    Let $\theta=\log(x)$ and $F=\mathbb{Q}(x)(\log(x))$. Let 
    \begin{center}
        $b=\theta^3 + 2x\theta^2 + x^2\theta + \theta^2 + 2x\theta + x^2$.
    \end{center} 
    We wish to find all $R \in F$ such that $\frac{R}{b}$ is elementary integrable.
As with Theorem \ref{theorem:Hermite}, we first compute the square-free factorization of $b$ to find $b=(\theta+1)(\theta+x)^2$. The partial fraction representation in this case will be 
\begin{center}
    $
   \frac{R}{b} = \frac{a(x)}{(\theta+1)} + \frac{b(x)}{(\theta+x)} + \frac{c(x)}{(\theta+x)^2}$
\end{center}
and we wish to find $a, b, c \in F_{n-1}$ that makes the entire expression elementary integrable. Since $b$ is not square-free, one iteration of Hermite Reduction is done to produce:
\begin{align*}
    &\int \frac{R}{b} =-\frac{c \! \left(x \right) x}{\left(1+x \right) \left(\theta +x \right)} \\
    &\quad + \bigintsss \frac{\left(a \! \left(x \right)+b \! \left(x \right)\right) \theta +a \! \left(x \right) x +b \! \left(x \right)+\left(\frac{\left(\frac{d}{d x}c \left(x \right)\right) x}{1+x}+\frac{c \left(x \right)}{1+x}-\frac{c \left(x \right) x}{\left(1+x \right)^{2}}\right) \left(\theta +1\right)}{\left(\theta +1\right) \left(\theta +x \right)}.
\end{align*}

Let us focus on the resulting integral: the denominator is $(\theta + 1)(\theta + x)$ which is now square-free. Thus, Hermite Reduction is no longer needed and instead, the TR-method is used on it. When the resultant is calculated and the roots of the TR-resultant are solved for (so that Theorem \ref{theorem:TR} is true), we get that the distinct roots are:
\begin{center}
    $ \left\{x a \! \left(x \right), 
\frac{x \left(\left(\frac{d}{d x}c \! \left(x \right)\right) x^{2}+b \! \left(x \right) x^{2}+\left(\frac{d}{d x}c \! \left(x \right)\right) x +2 b \! \left(x \right) x +b \! \left(x \right)+c \! \left(x \right)\right)}{x^{3}+3 x^{2}+3 x +1}.
\right\}$
\end{center}
Setting the first root to a constant is trivial to solve: $a(x)=\frac{C_1}{x}, C_1 \in \mathbb{Q}$. The second root condition contains the unknowns $b(x)$ and $c(x)$. This can also be set equal to a constant and then solved for $b(x)$ obtaining
\begin{center}
    $b \! \left(x \right) = 
\frac{-\left(\frac{d}{d x}c \! \left(x \right)\right) x^{2}-c \! \left(x \right) x +C_2}{x}, C_2 \in \mathbb{Q}$.
\end{center}

This means $c(x)$ can be any function from $F_{n-1}$. Let us demonstrate this by trying some values that are arbitrarily chosen:
\begin{itemize}
    \item[$\bullet$] $C_1=2 \implies a(x)=\frac{2}{x}$
    \item[$\bullet$]  $C_2=4$ and $c(x)=x^2 + \frac{1}{5x} \implies b(x)=\frac{-10 x^{4}+5 x^{3}+60 x^{2}+61 x +20}{5 x \left(1+x \right)^{2}}$
    \item[$\bullet$]  $\frac{R}{b} = \frac{2}{x \left(\ln \! \left(x \right)+1\right)}+\frac{-10 x^{4}+5 x^{3}+60 x^{2}+61 x +20}{5 x \left(1+x \right)^{2} \left(\ln \! \left(x \right)+x \right)}+\frac{x^{2}+\frac{1}{5 x}}{\left(\ln \! \left(x \right)+x \right)^{2}}$
    \item[$\bullet$] Then when we integrate $\frac{R}{b}$, we get:
\begin{center}
    $\int \frac{R}{b} = -\frac{5 x^{3}+1}{5 \left(1+x \right) \left(\ln \! \left(x \right)+x \right)}+2 \ln \! \left(\ln \! \left(x \right)+1\right)+4 \ln \! \left(\ln \! \left(x \right)+x \right)$
\end{center}
\end{itemize}
\end{example}
Example \ref{eg:hermite} gives us a much stronger freedom of choice because unlike with the square-free case, we actually get that our coefficient $c(x)$ can be \textit{any} function in $F_{n-1}$. This effectively means that we have three choices of freedom: one for $a(x)$ (the choice of the constant $C_1$), one for $b(x)$ (the choice of $C_2$), and one for $c(x)$ (any expression in the previous field). In contrast, the only choices of freedom we had in the square-free case were the constants. Additionally, Example \ref{eg:hermite} had one functional degree of freedom $c(x)$ since one factor from the denominator $b$ was quadratic. In general, we will have more functional degrees of freedom for higher degree factors in the denominator.  

\section{Discussion} \label{sec:Discussion}

The Risch algorithm is an integral part of any CAS (pun intended). This data generation method discusses how to create expressions that are guaranteed to be elementary integrable by using the Risch algorithm. To understand the benefit of this data generation method, we create a simple dataset of 10,000 (integrand, integral) pairs. To compare against our dataset, we take a sample of 10,000 data points from each of the FWD, BWD, and IBP datasets. Of the 10,000 we created, a third comes from generating polynomial expressions in Section \ref{sec:poly_gen}, another third comes from generating rational expressions from Section \ref{sec:rational_gen}, and the final third comes from combining the two sections together (similar to how the Risch algorithm separates the two parts from each other).  

\subsection{Risch Data Generation Benefits}

\begin{figure}
     \centering
     \begin{subfigure}[t]{0.48\textwidth}
         \centering
         \includegraphics[width=0.95\textwidth]{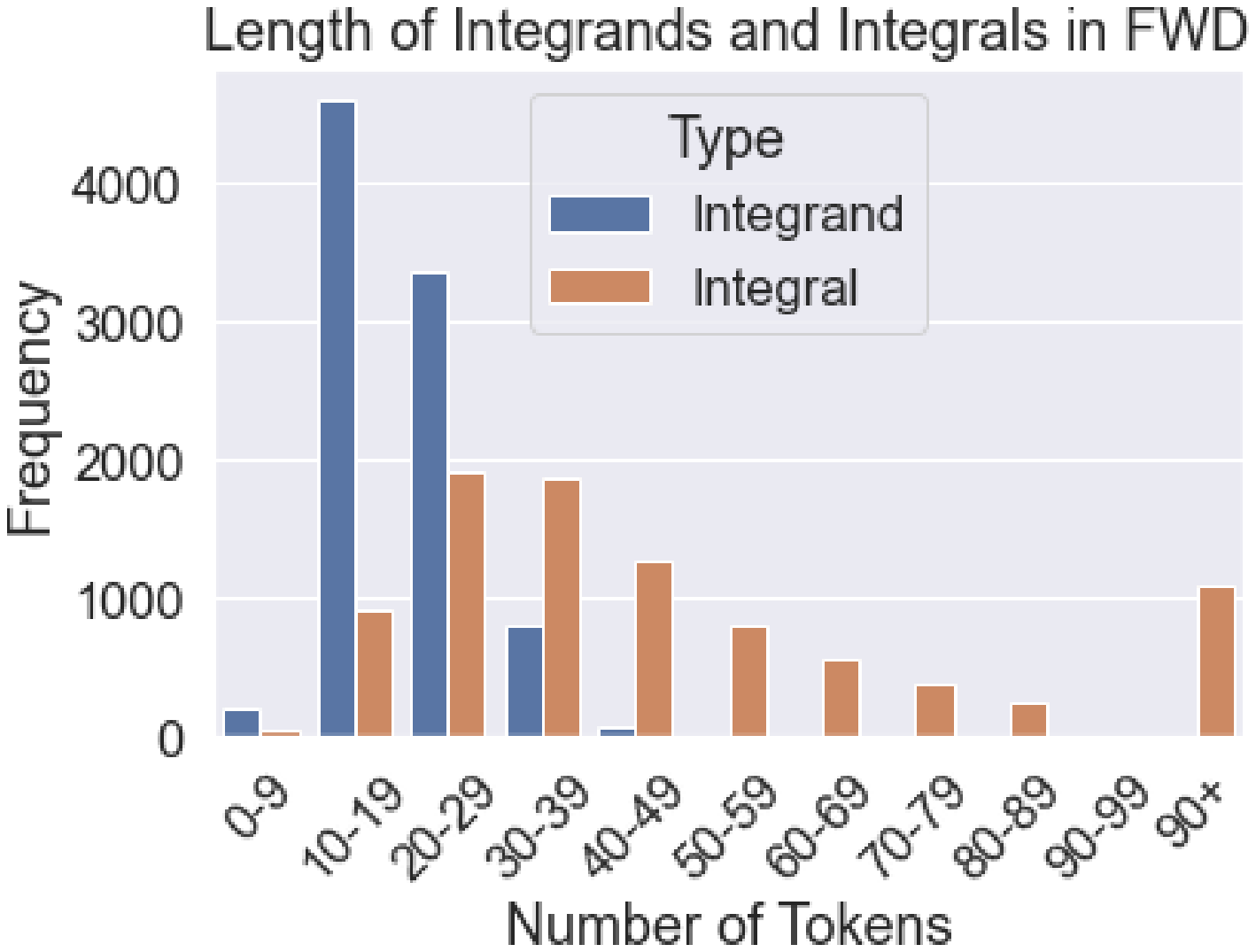}
         \caption{FWD}
         \label{fig:FWD_len}
     \end{subfigure}
     \hfill
     \begin{subfigure}[t]{0.48\textwidth}
         \centering
         \includegraphics[width=0.95\textwidth]{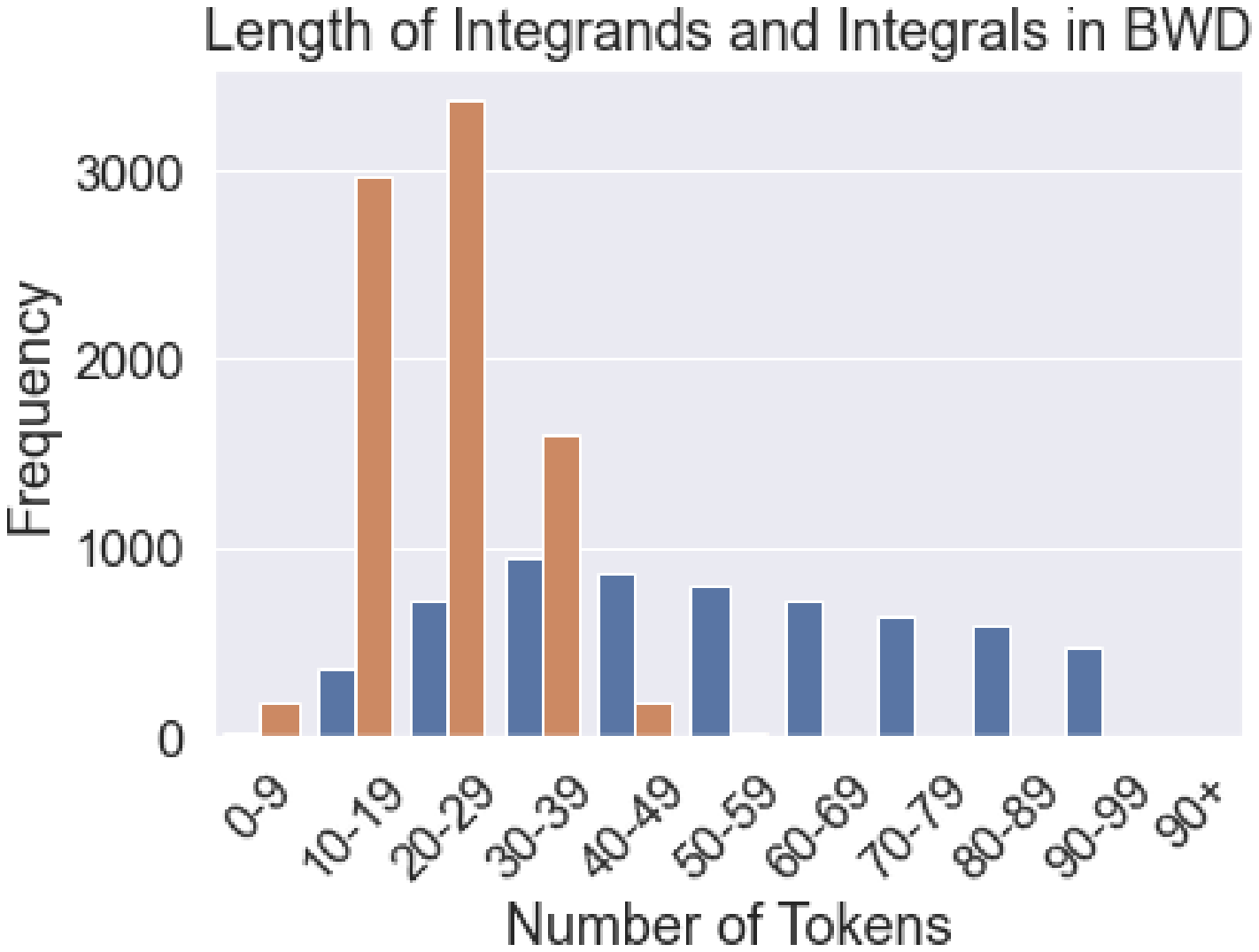}
         \caption{BWD}
         \label{fig:BWD_len}
     \end{subfigure}
     \hfill
     \vspace{0.6cm}
     \begin{subfigure}[b]{0.48\textwidth}
         \centering
         \includegraphics[width=0.95\textwidth]{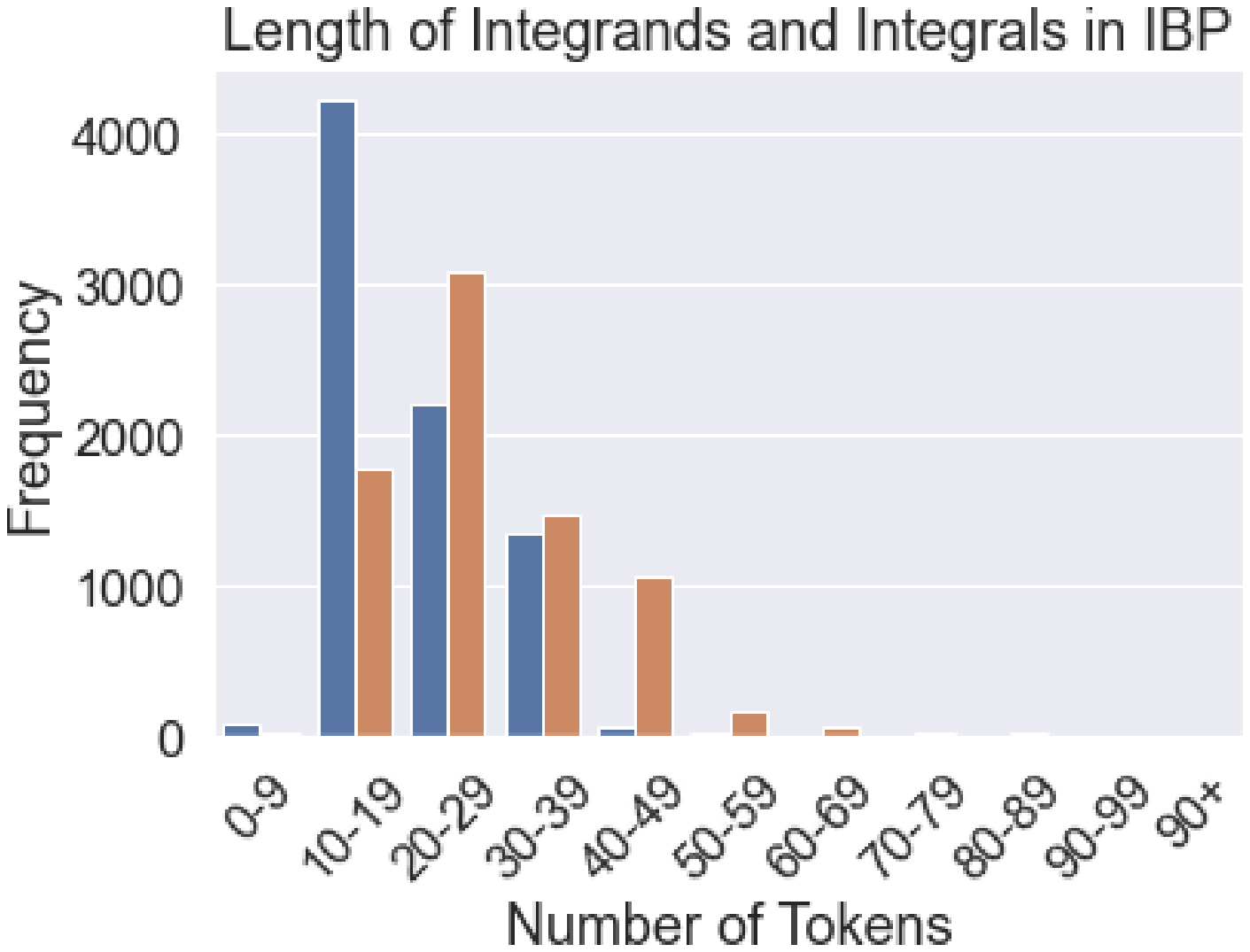}
         \caption{IBP}
         \label{fig:IBP_len}
     \end{subfigure}
     \hfill
     \begin{subfigure}{0.48\textwidth}
         \centering
         \includegraphics[width=0.95\textwidth]{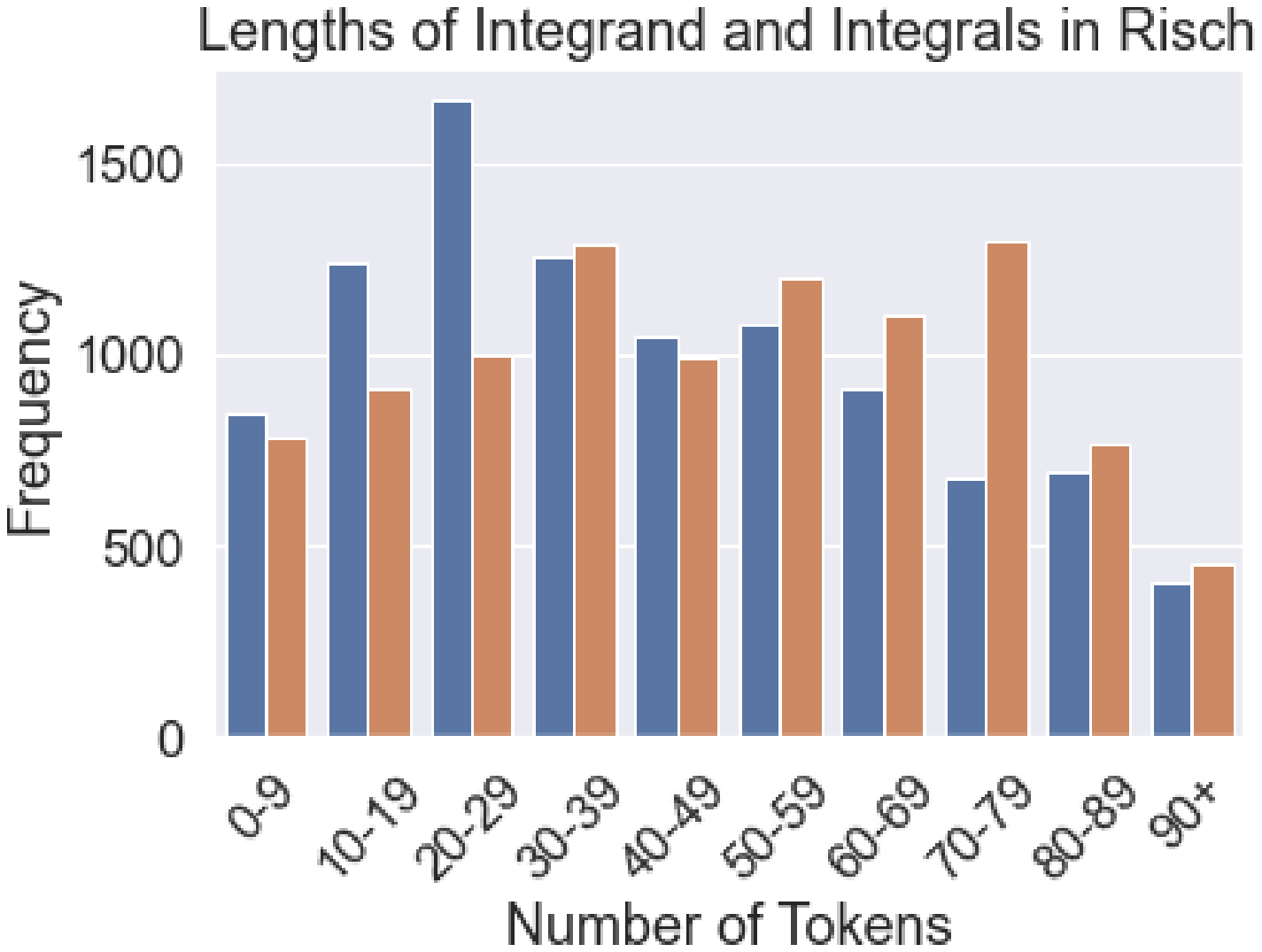}
         \caption{Risch}
         \label{fig:Risch_len}  
     \end{subfigure}
        \caption{Lengths of the Integrands and Integrals from the three test datasets in \cite{Lample2020} as well as our generated dataset.}
        \label{fig:LC_method_lens}
\end{figure}

One criticism of the data generation method in \cite{Lample2020} was that there were patterns within how the expressions are made, specifically in the FWD and BWD datasets. Recall from Section \ref{sec:current_methods}, the BWD method produced long integrands and short integrals whereas the FWD had the opposite problem. We take a closer look by examining the lengths of the integrands and integrals in their testing datasets. Note that the authors represent the mathematical expression in prefix (or normal Polish) notation. The length is then just the number of tokens from this representation. The lengths of the (integrand, integral) pairs are shown for all three data methods in Figure \ref{fig:LC_method_lens}.

Based on Figure \ref{fig:LC_method_lens}, we can see quite the difference in lengths from the FWD and BWD methods. Suppose we consider an (integrand, integral) pair close in length if the absolute value between the length of the integrand and integral is less than 10. For the FWD and BWD methods, only 29\% and 9\% of pairs were considered close respectively. The IBP and Risch methods do considerably better at generating close pairs with 65\% and 86\% of pairs being considered close respectively. As mentioned earlier in Section~\ref{sec:current_methods}, the presence of these patterns mean that there is a risk of bias in an ML model trained on such data.  Recall also from Section~\ref{sec:current_methods} how much of the data only differed by the choice of constants in the expression, making IBP a weaker generation method.

However, because of the choices of freedom we have in making our integrable expressions from the Risch algorithm, we can alleviate the two problems shown. This is true for both the polynomial expressions, the rational expressions, and a combination of the two. The only patterns present in our dataset are those required for the expression to be elementary integrable.  

With the dataset generated, Figure \ref{fig:Risch_len} shows the lengths of the produced (integrand, integral) pairs through the Risch algorithm in prefix notation. Figure \ref{fig:Risch_len} shows that the lengths between the integrands and integrals are much more evenly distributed, fixing the problem of the FWD and BWD datasets. Recall that the FWD method is also not able to generate (integrand, integral pairs) often, leading to a slow data generation method. Our method guarantees integrands that are elementary integrable 100\% of the time, making it more efficient. Furthermore, we do the same analysis of examining the dataset by substituting the integer coefficients with a \texttt{CONST} token, and find that 97\% of the data remains unique. The reason it did not reach 100\% is due to data generated in Section \ref{sec:TR_gen}, the rational square-free case. The choices of freedom in this case is usually only the choice of the constant. Some randomly generated denominators happened to be the same through chance and since the solutions only differ by a constant, they end up being the same when replaced with a \texttt{CONST} token. If wanted, these can be removed from the dataset.

\subsection{Future Work}

We have presented a novel method of creating elementary integrable functions.  However, there is much work that could still be done. Bronstein \cite{Bronstein2005}, when first introducing the Risch algorithm, separates the algorithm into four different cases: logarithmic transcendental, exponential transcendental, pure algebraic and mixed algebraic / transcendental cases. So far, we have only explored the first two cases. It would be beneficial to understand the latter two cases as radicals are something that should not be excluded from the dataset. To understand the latter two cases, one can read \cite{Bronstein1990} or \cite{Trager1984}. As with the present paper, the idea would be to find the conditions in the polynomial and rational cases that make the entire expression elementary integrable.

Furthermore, the current data generation method proposed can be further explored in a number of ways. For one, towers of extensions (i.e. $F_n, n \geq 2$) have only been considered for polynomial expressions thus far. This can also be done with the rational expression generation method to create a greater variety of elementary integrable expressions. Also, working with irreducible cubic and higher degree polynomials (in $\theta$) for the rational case should further be examined. We have shown that when we assume the form of the numerator (Example \ref{eg:rational_hard}), we can find solutions. However, it would be desirable to find \textit{all} possible numerators that make the entire expression integrable. The key to this would be examining the TR-resultant and instead of explicitly solving for the roots, qualitatively analysing the resultant and figuring out the conditions of the generic coefficients would help overcome the computational cost of explicitly solving for the solution as discussed at the end of Section \ref{sec:TR_gen}.

\subsubsection{Acknowledgements:} The authors would like to thank James Davenport and Gregory Sankaran for helpful discussion on conditions around constant roots of polynomials.  They would also like to thank John May for help understanding Maple's integration command and testing data and the anonymous reviewers for their comments which improved the paper.

Matthew England is supported by EPSRC Project EP/T015748/1, \emph{Pushing Back the Doubly-Exponential Wall of Cylindrical Algebraic Decomposition} (DEWCAD).  Rashid Barket is supported on a scholarship provided by Maplesoft and Coventry University.

%
%
%
\bibliographystyle{splncs04}
\bibliography{ref.bib}

\end{document}